\documentstyle[multicol,aps,prl,floats,epsf]{revtex}
\begin{document}
\renewcommand{\thesection}{\arabic{section}}

\draft

\title{
Non-universality of the scaling exponents\\
of a passive scalar convected by a random flow}
\author{M. Chertkov$^a$, G. Falkovich$^a$ and V. Lebedev$^{a,b}$}
\address{$^a$ Physics of Complex Systems, Weizmann Institute of Science,
Rehovot 76100, Israel \\ $^{b}$ Landau Institute for Theoretical 
Physics, Moscow, Kosygina 2, 117940, Russia}
\maketitle

\begin{abstract}
We consider passive scalar convected by multi-scale random velocity 
field with short yet finite temporal correlations. Taking Kraichnan's 
limit of a white Gaussian velocity as a zero approximation we develop perturbation 
theory with respect to a small correlation time and small
non-Gaussianity of the velocity. We derive the renormalization (due
to temporal correlations and non-Gaussianity) of the operator
of turbulent diffusion. That allows us to calculate the respective corrections
to the anomalous scaling exponents of the scalar field  and show that they 
continuously depend 
on velocity correlation time and the degree of non-Gaussianity. The scalar
exponents are thus non universal as was predicted by Shraiman and Siggia on
a phenomenological ground (CRAS {\bf 321}, 279, 1995). 
\end{abstract}


\begin{multicols}{2}

The most striking feature of turbulence is its intermittent spatial and 
temporal behavior. Statistically, intermittency means substantial
non-Gaussianity. For developed turbulence, where
the correlation functions are scale invariant at the inertial interval of 
scales, the intermittency is manifested as an anomalous 
scaling of correlation functions. That means that 
some random field $\theta({\bf r},t)$ has the structure functions 
$S_{2n}\!=\!\langle [\theta(t,{\bf r}_1)\!-\!\theta(t,{\bf r}_2)]^{2n}\rangle
\propto r_{12}^{\zeta_{2n}}$ with the exponents 
$\zeta_{2n}$ that are not equal to $n\zeta_2$. As a result, the degree 
of non-Gaussianity, that may be characterized by the 
ratio $S_{2n}/S_2^n$, depends on scale. Experiments and simulations show that 
the intermittency and anomalous scaling of the scalar field passively 
convected by a fluid are much stronger pronounced than the intermittency
of the velocity field itself \cite{84AHGA,91Sre,94HS,95KYC}. It is in the
problem of passive scalar where consistent analytic theory of an anomalous 
scaling starts to appear \cite{95KYC,68Kra-a,95GK,95CFKLb,95SS}.
It is intuitively clear that the physical reason for scalar intermittency
is a spatial inhomogeneity of the advecting velocity. The analysis of the
velocity field with smooth inhomogeneity shows, however, 
that there is no anomalous scaling of the scalar whatever be the 
(finite) temporal correlations of the velocity \cite{68Kra-a,94SS,95CFKLa}. 
Analytic treatment of a non-smooth multi-scale velocity
was possible hitherto only in the so-called Kraichnan's problem of white
advected scalar \cite{68Kra-a} where the correlation functions satisfy closed
linear equations of the second order \cite{94SS}. 
It has been shown \cite{95GK,95CFKLb} that, even without
any temporal correlations, spatial non-smoothness of the velocity 
provides for an anomalous scaling of the scalar. The anomalous parts
appeared as zero modes of the operator of turbulent diffusion and entered 
the correlation functions due to matching conditions at the pumping scale 
\cite{95GK,95CFKLb,95SS}. The coefficients
at the modes were thus pumping-dependent while the form of any zero
mode was universal i.e. determined only by the exponent of the 
velocity spectrum and space dimensionality. In particular, 
the exponents $\zeta_n$ of the scalar were universal for the delta-correlated
velocity.

Now, what is the role of velocity temporal behavior in building up 
intermittency of the scalar field? It was argued phenomenologically 
by Shraiman and Siggia \cite{95SS} that the exponents of the scalar 
field depend on more details of the velocity statistics ``than just 
exponents''. Here, we consider the simplest possible generalization 
of Kraichnan's problem and consistently derive the equations for 
scalar correlation functions in the case of short yet finite 
velocity correlation time $\tau_r$ which is supposed to be a power 
function of the scale $r$. The behavior of the ratio $\tau_r/t_r$ 
is important, where $t_r$ is the turnover time
at the scale $r$. If the ratio tends to zero at decreasing $r$ then we
return to the $\delta$-correlated case. If the ratio increases at
decreasing $r$ we encounter the problem of the quenched disorder type
which should be considered separately. We consider the marginal
case of a complete self-similarity
where $\epsilon=\tau_r/t_r$ does not depend of $r$ and
formulate the perturbation theory regarding the ratio as the small 
parameter of our theory. We show that $\zeta_2$ does not depend on 
$\epsilon$ while $\zeta_n$ for $n>2$ are $\epsilon$-dependent  
that is the set of the exponents is non universal along with
the prediction of \cite{95SS}. The principal difference 
between the second and higher correlation functions is naturally
explained on the language of zero modes: there is no zero mode (except 
constant) for the pair correlator while the zero modes of the high 
correlators depend on the precise form of the operator of turbulent 
diffusion which is $\epsilon$-dependent. This is formally similar to what 
has been discovered by Kadanoff, Wegner and Polyakov in the theory 
of phase transitions: the critical exponents continuously 
depend on the amplitude of the operator term with dimension $d$
added to the Hamiltonian \cite{71KW,73Pol}. 

Note that the below results cannot be directly applied to the description 
of scalar advection by a Kolmogorov turbulence: because of sweeping effect, 
the different-time velocity statistics is not scale-invariant in the 
Eulerian frame \cite{64Kr}. Our use 
of scale-invariant velocity is intended to establish the general fact
of the sensitive dependence of scalar exponents on the velocity statistics.

The advection of passive scalar $\theta(t,{\bf r})$ by an incompressible 
flow is governed by the equations
\begin{equation}
(\partial_t\!-\!\hat{P})\theta\!=\!\phi,\quad 
\hat{P}(t)\!=\!-\!v^\alpha\nabla^\alpha
\!+\!\kappa\triangle,\quad \nabla^\alpha v^\alpha\!=\!0,
\label{em} \end{equation}
where $\kappa$ is the coefficient of molecular  diffusion. The
advecting velocity ${\bf v}$ and the source $\phi$ are independent
random functions. A formal solution of (\ref{em}) is
\begin{eqnarray} && 
\theta(t,{\bf r})=\int_{-\infty}^t dt_1 
{\sl T}\exp\biggl(\int_{t_1}^{t} dt'\, 
\hat{P}(t')\biggr)\phi(t_1,{\bf r}) \,,
\label{a2a} \end{eqnarray}
where ${\sl T}\exp$ designates the chronologically ordered exponent. From 
(\ref{a2a}) it follows 
\begin{eqnarray} && 
F_n(t,{\bf r}_1,\dots,{\bf r}_{2n})\equiv
\langle\theta(t,{\bf r}_1)\dots\theta(t,{\bf r}_{2n})\rangle
\nonumber \\ &&
=\int\limits_{-\infty}^t\!\!\! dt_1 \dots
\int\limits_{-\infty}^t\!\!\! dt_{2n}
\hat{\cal A}\, \Biggl\langle \prod_{i=1}^{2n} 
\phi(t_i,{\bf r}_i)\Biggr\rangle \,,
\label{co} \\ && 
\hat{\cal A}=\langle \hat Q \rangle \,, \quad
\hat Q(t)=\prod_{i=1}^{2n}{\sl T}\exp 
\left(\int_{t_i}^t dt'_i\hat P(t'_i,{\bf r}_i)\right) \,.
\label{ti1} \end{eqnarray}
Differentiating $\hat{\cal A}$ over the current time $t$, one gets 
\begin{equation}
\partial_t{\hat{\cal A}}=
\left\langle\hat{\cal P}(t)\hat Q(t)\right\rangle \,, \quad
\hat{\cal P}=\kappa\nabla_i^2-v_i^\alpha\nabla_i^\alpha \,,
\label{ti2} \end{equation}
where ${\bf v}_i={\bf v}(t,{\bf r}_i)$ and 
$\nabla_i=\partial/\partial{\bf r}_i$. 
Here and below summation over both repeated vector superscripts and subscripts
enumerating points ${\bf r}_i$ is implied. The identity (\ref{ti2}) 
can be brought to the form
\begin{equation}
\partial_t\hat{\cal A}(t)
=\kappa\nabla^2_i\hat{\cal A}(t)
+\int\limits_0^\infty dt' 
\hat{\cal N}(t')\hat{\cal A}(t-t') \,,
\label{co1} \end{equation} 
where $\hat{\cal N}(t)$ is to be found. The decay of $\hat{\cal N}(t)$
is determined by the velocity correlation time $\tau_r$ which is supposed 
to be much smaller than the spectral transfer time characteristic of
(\ref{co}). It is the reason why the upper limit in (\ref{co1}) can
be substituted by infinity.

We shall find the first terms of the expansion of $\hat{\cal N}$ in  $\tau_r$. 
Let us first examine the Gaussian contribution to $\partial_t\hat{\cal A}$ 
related to reducible correlation functions of ${\bf v}$
\begin{equation}
\int^t d\tilde t\, \underline{v}_i^\alpha(t)\nabla_i^\alpha
\bigg\langle{\sl T}\exp
\Big[\int\limits_{\tilde t}^t dt'
\hat{\cal P}(t')\Big] \underline{v}_j^\beta(\tilde t)\nabla^\beta_j
\hat Q(\tilde t)\bigg\rangle_{\rm G} ,
\label{ti3} \end{equation}
where the product $\underline{v}_i^\alpha(t)\underline{v}_j^\beta(\tilde t)$
should be substituted by the pair correlation function
$\langle v^\alpha(t,{\bf r}_i)v^\beta(\tilde t,{\bf r}_j)\rangle$.
The integrand is nonzero for $t-\tilde t\leq\tau_r$
and consequently ${\sl T}\exp
\Big[\int_{\tilde t}^t dt'\hat{\cal P}(t')\Big]$ can be expanded over 
$t-\tilde t$. The zero term gives
\begin{equation}
\hat{\cal N}_0(t)=\langle v^\alpha_i(t)\nabla^\alpha_i
v^\beta_j(0)\nabla^\beta_j \rangle \,.
\label{co2} \end{equation}
The first and the second terms of the expansion 
produce the linear in $\tau_r$ contribution 
\begin{eqnarray} && 
\hat{\cal N}_1(t)\!=\!
\int\limits_0^t\! dt_1\biggl[\int\limits_0^{t_1}\! dt_2
[\underline{v}_i^\alpha(t)\nabla_i^\alpha
\overline v_k^\gamma(t_1)\nabla_k^\gamma
\underline{v}_j^\beta(t_2)\nabla^\beta_j
\overline v_m^\mu(0)\nabla_m^\mu 
\nonumber \\ &&
+\underline{v}_i^\alpha(t)\nabla_i^\alpha
\overline v_k^\gamma(t_1)\nabla_k^\gamma
\overline v_m^\mu(t_2)\nabla_m^\mu
\underline{v}_j^\beta(0)\nabla^\beta_j]
\label{nz} \\ &&
+\kappa t_1\underline{v}_i^\alpha(t_1)\nabla_i^\alpha\nabla_k^2
\underline{v}_j^\beta(0)\nabla_j^\beta\biggr],
\nonumber \end{eqnarray}
where the products $\overline v\,\overline v$ and $\underline v\,\underline v$ 
should be substituted by the corresponding pair correlation functions. 

For a short-correlated velocity field, the leading non-Gaussian
contribution to the correlation functions of ${\bf v}$ is determined by 
the irreducible part of the fourth-order correlation function of
${\bf v}$. Generalizing the trick leading from (\ref{ti2}) to (\ref{ti3}) 
we obtain the non-Gaussian term $\hat{\cal N}_{\rm nG}(t)$:
\begin{equation}
\!\!\int_0^t\!\!dt_1\!\!\int_0^{t_1}\!\!d t_2
\langle\!\langle\! v_i^\alpha(t)\nabla_i^\alpha v_k^\mu(t_1)\nabla_k^\mu
v_j^\beta(t_2)\nabla_j^\beta 
v_n^\gamma(0)\nabla_n^\gamma 
\!\rangle\!\rangle\,,
\label{nog} \end{equation}
where double angular brackets stand for the cumulant. 

The operator $\hat{\cal A}(t)$ is exponential in time
\begin{equation}
\hat{\cal A}(t)=\exp\bigl[(t-t_0)(\kappa\nabla^2_i
+\hat{\cal L})\bigr]\hat{\cal A}(t_0)
\label{th} \end{equation}
asymptotically at $t-t_0\gg\tau_r$.
Substituting (\ref{th}) into (\ref{co1}), expanding $\exp(t'\hat{\cal L})$
and keeping only the principal terms we find the operator of turbulent diffusion
\begin{eqnarray} &&
\hat{\cal L}=\hat{\cal L}_0+\hat{\cal L}_1
+\hat{\cal L}'_1+\hat{\cal L}_{nG}\, 
\label{tj2} \\ &&
\hat{\cal L}_{\{0,1,nG\}}\!
=\!\!\int_0^\infty\!dt\,\hat {\cal N}_{\{0,1,nG\}}(t),\, 
\ \hat{\cal L}'_1\!=\!\!-\!\!
\int_0^\infty\!dt\, t {\cal N}_0(t)\hat{\cal L}_0 \,.
\nonumber \end{eqnarray}
Using (\ref{co},\ref{th}) we can obtain the expression for
$\partial_t F_n$. For the pumping $\delta$-correlated in time, one gets
\begin{eqnarray} &&
\partial_t F_n(t,{\bf r}_1,\dots,{\bf r}_{2n})
-\hat{\cal L}F_n(t,{\bf r}_1,\dots,{\bf r}_{2n})
\nonumber \\ &&
=\hat{\cal M}[\chi_{12}F_{n-1}(t,{\bf r}_3,\dots,
{\bf r}_{2n})+{\rm permutations}] \,.
\label{eq} \end{eqnarray}
Here, the function 
$\chi(r_{12})=\int dt\,\langle\phi(t,{\bf r}_1)\phi(0,{\bf r}_2)\rangle$
decays on the pumping scale $L$ and $\chi(0)$ is the production 
rate of $\theta^2$. The operator $\hat{\cal M}$ in (\ref{eq}) 
can be estimated as $\hat{\cal A}(\tau_L)$. The
account of temporal correlations of the pumping (which can be done
perturbatively as long as the pumping correlation time is much less than
the time of scalar transfer) results in an extra renormalization of 
$\hat{\cal M}$ operator. Its explicit form is unimportant 
for what follows. Indeed, the balance equation
(\ref{eq}) contains the renormalization (due to velocity temporal correlations
and non-Gaussianity) of all three relevant quantities: pumping, turbulent
diffusion and molecular diffusion (the last term in  $\hat {\cal N}_1$).
We discuss here only the scaling exponents in the convective
interval of scales (see below) that are determined solely by the form of
the operator of turbulent diffusion $\hat{\cal L}$.

Let us consider the pair correlation function of the velocity to 
be scale-invariant:
\begin{eqnarray}&& 
\langle[v^\alpha(t,{\bf r})\!-\!v^\alpha(0,{\bf 0})]
[v^\beta(t,{\bf r})\!-\!v^\beta(0,{\bf 0})]\rangle\!=\!2
K^{\alpha\beta}(t,r),
\nonumber \\ &&
K^{\alpha\beta}\!=\!\frac{D r^{2-\gamma}}{\tau_r}\biggl[
\biggl(\!\delta^{\alpha\beta}\!
-\!\frac{r^\alpha r^\beta}{r^2}\!\biggr)g_\bot
\biggl(\!{|t|\over\tau_r}\!\biggr)\!+\!
\delta^{\alpha\beta}g_\|\biggl(\!{|t|\over\tau_r}\!\biggr)\biggr]
\label{vel} \end{eqnarray}
with the correlation time $\tau_r\!=\!\tau_L (r/L)^z$.
Dimensionless functions $g_\bot$ and $g_\|$ satisfy the 
incompressibility condition
$(d\!-\!1)g_\bot(x)\!=\!zx^{a}d[x^{1-a}g_\|(x)]/dx$ where
$a\!=\!({2\!-\!\gamma})/{z}-1$. Their normalization is fixed by below
expressions (\ref{ko},\ref{kw}). The main term in (\ref{tj2}) is \cite{94SS}
\begin{eqnarray} &&
\hat{\cal L}_0=-\sum_{ij}
{\cal K}_0^{\alpha\beta}(r_{ij})\nabla_i^\alpha\nabla_j^\beta \,
\quad {\cal K}_0^{\alpha\beta}
=2\int\limits_0^\infty dt\,{\cal K}^{\alpha\beta}(t)\,
\label{lo} \\ &&
{\cal K}_0^{\alpha\beta}=Dr^{2-\gamma}
\left(\frac{d+1-\gamma}{2-\gamma}\delta^{\alpha\beta}
-\frac{r^\alpha r^\beta}{r^2}\right) \,.
\label{ko} \end{eqnarray}
The expressions (\ref{lo},\ref{ko}) lead to the following
turnover time $t_r\!=\!(2\!-\!\gamma)r^\gamma/D\gamma d(d\!-\!1)$ obtained 
for the delta-correlated case \cite{68Kra-a,95CFKLb}. Our
marginal case corresponds to $z=\gamma$ and the small parameter of the 
perturbation theory is thus
\begin{equation}
\epsilon={D\tau_L}L^{-\gamma}{\gamma d(d-1)}/{(2-\gamma)}\ll 1.
\label{c4} \end{equation}
Note that $\epsilon$ contains $d^2$
which tells us that the space dimensionality should not
be very large for the approximation of a short correlation to be valid: 
the characteristic time of the scalar transfer (proportional to $d^{-2}$) 
should be larger than the correlation time. 

Starting from the expression for the pair velocity correlator (\ref{vel})
we can obtain the first Gaussian $\epsilon$-correction to (\ref{lo}).
Calculating (\ref{nz}) and then integrals in (\ref{tj2}) we find 
\end{multicols}
\begin{eqnarray} &&
\hat{\cal L}_1\!+\!\hat{\cal L}_1^\prime\!=\!
\frac{1}{2}\sum_{i,j,k}K_{0;ij}^{\alpha\beta}
K_{1;ik}^{\mu\nu;\alpha}
\nabla_i^{\mu}\nabla_j^{\beta}\nabla_k^{\nu}\!-\!
\frac{1}{2}\sum_{i,j}{\cal B}_{ij}^{\mu\nu}
\nabla_i^{\mu}\nabla_j^{\nu}\!-\!
\frac{\kappa}{2}\sum_{i,j,k}\nabla^2_k
{\cal K}_{1;ij}^{\alpha\beta}\nabla_i^\alpha \nabla_j^\beta,
\quad
{\cal B}^{\mu\nu}({\bf r})\!=
\!K_{1;ij}^{\alpha\mu;\beta}
K_{0;ij}^{\beta\nu;\alpha}\!\!
-\!K_{1;ij}^{\alpha\beta}
K_{0;ij}^{\mu\nu;\alpha\beta}\!\!
\nonumber \\ &&
+ 2\!\int_0^\infty\!\! dt_1\nabla_{\bf r}^{\alpha}
\nabla_{\bf r}^{\beta}\biggr[
\int_{t_1}^\infty\!\!\!
dt_2 K^{\alpha\beta}(t_2;{\bf r})
\int_{t_1}^\infty\!\!\!
dt_3 K^{\mu\nu}(t_3;{\bf r})
-\int_{t_1}^\infty\!\!\!
dt_2 K^{\alpha\mu}(t_2;{\bf r})
\int_{t_1}^\infty\!\!\!
dt_3 K^{\beta\nu}(t_3;{\bf r})\biggr],
\label{c9}\end{eqnarray}
\begin{multicols}{2}
\noindent
where $K^{\alpha\beta;\mu}_n\equiv\nabla_{\bf r}^{\mu} K^{\alpha\beta}_n$,
$K^{\alpha\beta;\mu\nu}_n\equiv \nabla_{\bf r}^{\mu} \nabla_{\bf r}^{\nu} 
K^{\alpha\beta}_n$ and
\begin{eqnarray} &&
K_1^{\alpha\beta}\!({\bf r})\!=\!
2\!\int\limits_0^\infty\! dt 
K^{\alpha\beta}(t,{\bf r})\!=\!Dr^{2-\gamma}\tau_r
\left(\!\frac{d\!+\!1}{2}\delta^{\alpha\beta}\!-
\!\frac{r^\alpha r^\beta}{r^2}\right)\,,
\nonumber \\&&
{\cal B}^{\alpha\beta}({\bf r})=\epsilon D r^{2-\gamma}\left[
b_\|\delta^{\alpha\beta}+b_\bot\left(\delta^{\alpha\beta}-
\frac{r^\alpha r^\beta}{r^2}\right)\right]\ .
\label{kw} \end{eqnarray}

Now we can analyze the equation (\ref{eq}) for $F_n$.
At the convective interval of scales 
$L\!\gg\! r\!\gg\! [\kappa(2\!-\!\gamma)/D(d\!-\!1)]^{1/(2\!-\!\gamma)}$,
the molecular diffusion term can be dropped:
it is enough to require $F_n=0$ at $r=0$ \cite{95GK,95CFKLb}.
Here, the zero modes of $\hat{\cal L}$ are responsible 
for the anomalous scaling of $F_{n}$.  
The scaling exponents of the bare operator
$\hat{\cal L}_0$ and the perturbation operator $\hat{\cal L}_1+
\hat{\cal L}_1^\prime$ coincide. For a self-similar velocity statistics, the
non-Gaussian contribution $\hat{\cal L}_{nG}$ 
has the same scaling too. The first consequence is that the exponent of 
the pair correlation function is $\gamma$ at 
arbitrary finite order in $\epsilon$ for any $\gamma$ and $d$. 
Indeed, there is no zero mode of the two-point $\hat{\cal L}$ 
with a nonzero positive exponent that could provide an anomaly. 
Contrary, for $n>2$, the account of the $\epsilon$-contributions to 
the bare operator $\hat{\cal L}_0$ should produce obviously
$\epsilon$-dependent corrections to the exponents of zero modes and
consequently $\epsilon$-dependent anomalous scaling.

To illustrate the above conclusion about 
$\tau$-dependence of the scalar exponents, let us give an 
example where the calculation can be done explicitly. We consider a 
large dimensionality (the limit $\gamma\gg (2-\gamma)/d$
solved in \cite{95CFKLb,95CF} for $\tau=0$) 
while assuming that, in addition to (\ref{c4}),
$1/d\gg\epsilon$ (it will be seen below how the parameter $\gamma$ enters the
condition). The leading (in $d$) terms 
of the bare and the Gaussian perturbative operators in terms of relative 
distances $r_{ij}$ are as follows [multiplied by $(2\!-\!\gamma)/dD$]
\end{multicols}
\begin{eqnarray}&&\hat{\cal L}_{0,0}\!\!=\!\!
d\!\sum\limits_{i>j}\!r_{ij}^{1-\gamma}\partial_{r_{ij}},\ 
\hat{\cal L}_{0,1}\!=\!\!
 \sum_{i>j}r_{ij}^{1-\gamma}\!\bigl(r_{ij}\partial^2_{ij}\!-\!
\gamma\partial_{ij}\bigr)\!\!-\!\!
\frac{1}{2}\!\!\sum_{i,j,p,q}\!r_{ij}^{2-\gamma}\frac{{\bf r}_{ip}{\bf r}_{jq}}
{r_{ip}r_{jq}}\partial_{ip}\partial_{jq},\nonumber\\&& 
\hat{\cal L}_{1,0}= {\epsilon d(2\!-\!\gamma)\over\gamma}\!
\left[\sum_{k>l}\!r_{kl}\partial_{kl}\!
+\!\gamma\!-\!1\!+\!\!2b_\|^{(0)}\!\right]\!
\sum_{i>j}\!\!r_{ij}^{1-\gamma}\partial_{ij},
\nonumber\\&&
\hat{\cal L}_{1,1}\! =\!
{\epsilon (\gamma-2)\over8\gamma}\biggl[\!\sum\!
\biggl(\frac{{\bf r}_{ip}{\bf r}_{jq}}{r_{ip}r_{jq}}
r_{ij}^2r_{kl}^{1-\gamma}\partial_{ip}\partial_{jq}\partial_{kl}\!+\!
\frac{{\bf r}_{kp}{\bf r}_{lq}}{r_{kp}r_{lq}}
r_{ij}r_{kl}^{2-\gamma}\partial_{kp}\partial_{lq}\partial_{ij}\biggr)\!
+\!4\!\sum\!\frac{{\bf r}_{jp}{\bf r}_{ik}}{r_{ik}r_{jp}}
\biggl(r_{ij}^{2-\gamma}\!+\!\frac{2-\gamma}{2}r_{ij}^2r_{ik}^{-\gamma}\biggr)
\partial_{jp}\partial_{ik}
\nonumber \\ && 
+4{b_\|^{(0)}}\!\sum\!
r_{ij}^{2-\gamma}\frac{{\bf r}_{ip}{\bf r}_{jq}}{r_{ip}r_{jq}}
\partial_{ip}\partial_{jq}\!+\!
16\!\sum_{i>j,k>l}\!r_{ij}^{1-\gamma}r_{kl}\partial_{ij}
\partial_{kl}\!+\!
\left(16\!-\!4\gamma\!-\!8\bigl(b_\|^{(1)}\!+\!b_\bot^{(0)}\!-\!
b_\|^{(0)}\bigr)\right)\!\sum_{i>j}\!
r_{ij}^{1-\gamma}\partial_{ij}\biggr],
\label{Lpp} \end{eqnarray}
\begin{multicols}{2}

\noindent
The summation is performed  over $n(n-1)/2$ distances which are independent
variables if $d\!>\!n\!-\!2$. For the chosen form of 
$K^{\alpha\beta}(t,{\bf r})$,
$b_\|\to (2-\gamma)( b_\|^{(0)}d^3+b_\|^{(1)}d^2)$, $b_\bot\to (2-\gamma)
b_\bot^{(0)}d^2$ at $d\to\infty$, with $d$-independent constants 
$b_{\|,\bot}^{(i)}$. First, we calculate the corrections to the exponents
related to the Gaussian correction (\ref{Lpp}) and then discuss the
corrections due to non-Gaussianity.
 
Solving the equation for the pair correlation function one can check that 
$\zeta_2\!=\!\gamma$ is independent of $\epsilon$ and $d$. Then, we consider 
the four-point correlation function ($i,j,k,l=1\ldots4$). To get the main 
contribution at $r\ll L$ one has to perturb the bare zero mode of 
$\hat{\cal L}_{0,0}+\hat{\cal L}_{0,1}$. In the limit under consideration,
it is enough to consider only the mode
\begin{equation}
Z_0=\!\sum_{\{i,j,k,l\}}\! (r^\gamma_{ij}\!-\!r^\gamma_{kl})^2
-1/2 \!\sum_{\{i,j,k\}}\! (r^\gamma_{ij}\!-\!r^\gamma_{ik})^2,
\label{zm1} \end{equation}
with the leading exponent $\Delta_4(0)=4(2-\gamma)/d$
found in \cite{95CFKLb} by $1/d$-expansion. The first
$\epsilon$-correction to (\ref{zm1}) can be obtained by applying the operator
$-\hat{\cal L}_{0,0}^{-1}
(\hat{\cal L}_{1,0}\hat{\cal L}_{0,0}^{-1}\hat{\cal L}_{0,1}
+\hat{\cal L}_{0,1}\hat{\cal L}_{0,0}^{-1}\hat{\cal L}_{1,0})$ to the mode 
$Z_0$. The correction to the anomalous exponent is determined by the
coefficient at $\ln(L/r)$ in the first $\epsilon$-contribution to $Z_0$, it is
\begin{equation}
\Delta_4(\epsilon)=\Delta_4(0)+{\epsilon(2-\gamma)\over d\gamma}
\bigl(4+6\gamma-2\gamma^2\bigr)\ .
\label{corr} \end{equation}
The sign of the correction is positive for 
$0\!<\!\gamma\!<\!2$. We can also calculate $\tau$-related
corrections for the high-order functions by using the technique
developed in \cite{95CF} for finding the largest anomalous exponents. 
For $n\!\ll\!\gamma d$,
\begin{equation}
\Delta_{2n}(\epsilon)={n(n-1)}\Delta_4(\epsilon)/2.
\label{ne} \end{equation}
The scaling exponents 
thus depend not only on purely dimensionless quantities $\gamma$ and $d$ yet
also on a dimensionless ratio of dimensional quantities. In other words, the
exponents depend on the form of the structural functions $g_\bot$ and $g_\|$.

Considering opposite hierarchy of small parameters $1/d\!\ll\!\epsilon $
and neglecting $1/d$ corrections, one finds
$\zeta_{2n}\!=\!n\zeta_2$ at any order in $\epsilon$:
temporal correlations by themselves do not produce an anomalous scaling if it 
is absent in the uncorrelated case. In
this limit, the anomalous exponents appear only in the next $1/d$ order
and are proportional to $\epsilon d$ -- see \cite{chert} for the details.

Now let us discuss the non-Gaussian contribution to the anomalous exponents.
We denote by $\epsilon_4$ the ratio of the cumulant to the fourth-order 
correlator and consider the limit $1\gg 1/d\gg\epsilon_4 d^3$.
Using the expressions (\ref{nog},\ref{tj2},\ref{eq}) we conclude that
the contribution to $\Delta_n$ is proportional to $n(n-1)\epsilon_4 d^2$
for $n\ll\gamma d$.

To conclude, we learned that the scalar exponents  are 
sensitive to the details of the velocity statistics. The existence of
two different contributions (due to temporal correlations and non-Gaussianity
of the velocity) makes it possible that there exist some classes of the
statistics with special relations between the contributions (some
remarkable cancelations, for instance), their analysis is left for future studies.
Hopefully, real turbulent flows belong to those classes and  analytic
expressions for the passive scalar exponents can be found some day.

We are grateful to E. Balkovsky, G. Eyink, I. Kolokolov and B. Shraiman
for useful discussions. This work was partly supported by the 
Clore Foundation (M.C.), by the Rashi Foundation (G.F.) 
and by the Minerva Center for Nonlinear Physics (V.L.).

\end{multicols}
\end{document}